\documentclass[14pt]{extarticle}
\usepackage[left=2cm,right=2cm]{geometry}
\usepackage{graphicx} % Required for inserting images
\usepackage{xcolor}
\usepackage{amssymb}
\usepackage{amsmath}
\usepackage{dsfont}
\usepackage{hyperref}
\usepackage{titlesec}
\usepackage{todonotes}
\usepackage{amsmath,amssymb,amsfonts,mathtools}
\usepackage{physics}
\usepackage{bm}
\usepackage[
    backend=biber,
    style=numeric, 
    sorting=none   
]{biblatex}
\usepackage{authblk}
\DeclareUnicodeCharacter{0300}{HEREHEREHERE}
\DeclareUnicodeCharacter{0304}{HEREHEREHERE}

\titleformat*{\section}{\bfseries}

\title{\Large A %classical-quantum 
one-world interpretation\\ of quantum mechanics}
\author[1]{Isaac Layton}
\author[2]{Jonathan Oppenheim}
\author[2]{Zachary Weller-Davies}
\affil[1]{Department of Applied Physics, The University of Tokyo,
7-3-1 Hongo, Bunkyo-ku, Tokyo 113-8656, Japan}
\affil[2]{Department of Physics and Astronomy, University College London, Gower Street, London WC1E 6BT, United Kingdom.}

\date{\large October 9, 2025}

\begin{document}

\maketitle

\begin{abstract}
The measurement problem is the issue of explaining how the objective classical world emerges from a quantum one. Here we take a different approach. We assume that there is an objective classical system, and then ask that the standard rules of probability theory apply to it when it interacts with a quantum system. %described in Hilbert space. 
Under mild assumptions, we recover the unitary dynamics, collapse and Born rule postulates from quantum theory. Nonetheless, there is no decoherence, because the quantum state remains pure conditioned on the classical trajectory. This results in one world, rather than many-worlds. Our main technical tool is to exploit a change of measure on the space of classical paths, the functional form of which is shown to characterise the quantum dynamics and Born rules of a class of quantum-like theories.
%, no distinction between unitary and collapse phases, contrast to collapse models the quantum state remains pure conditioned on the classical degrees of freedom,change of measure
    
\end{abstract}

\section{Introduction}

The measurement problem is the difficulty in adequately explaining the classical outcomes of measurements within the theory of unitary quantum mechanics. Rather than being an abstract problem, this leads to paradoxes in the theory if the rules of unitary quantum mechanics are naively followed \cite{frauchiger2018quantum}. This has led to a plethora of interpretations of quantum theory, including QBism \cite{Fuchs2014QBism}, decoherent histories \cite{halliwell1994review}, many worlds \cite{Everett1957RelativeState,Wallace2012EmergentMultiverse}, quantum darwinism \cite{Zurek2009QuantumDarwinism} and many others \cite{Bohm1952Interpretation,Spekkens2007ToyModel,Pusey2012PBR,Leifer2014IsQuantumStateReal,Ghirardi1986GRW,Bassi2003CollapseReview} to explain when and how classically definite observations arise.

%To address this, a number of interpretations have arisen... Attempting to use quantum theory to describe he understanding of how to apply interpretation of the theory may have

%To address this, a number of interpretations have arisen... Attempting to use quantum theory to describe he understanding of how to apply interpretation of the theory may have 

Here we circumvent this problem by postulating the existence \textit{a priori} of a %classical system. 
system which is treated classically.
We argue that this gives a cohesive interpretation of quantum theory, in which neither measurement nor sudden collapses of the wavefunction are necessary. Instead, quantum theory is reduced to a description of two interacting systems: one described using Hilbert space, the other using a classical configuration space.

 To support this interpretation, we show
 that with minimal assumptions,
%about classical probability theory
 the existence of a classical degree of freedom, along with classical probability theory, is sufficient to deduce much of the structure of quantum theory. In other words, one arrives at both the collapse postulate and the Born rule almost for free, once one postulates the existence of classical and quantum systems and a shared noise process. The resulting dynamics provide a simple and intuitive picture of quantum mechanics: classical and quantum systems evolve continuously in their respective configuration/Hilbert spaces; a single equation describes both the collapse and unitary phases of quantum evolution. At all times the quantum state remains pure, conditioned on %observations of 
 the classical system's trajectory. At all times the classical system is in a definite state, although its future evolution may be unpredictable. One need never speak of measurement, or collapse of the wavefunction: in this interpretation these merely become convenient concepts invoked to describe the backreaction of the quantum system on the classical one. Since both the classical and quantum state trace-out a single, definite trajectory, we say this provides a {\it one-world interpretation of quantum theory}\footnote{The term ``single-world'' has been used in other contexts, often to categorize non-Many Worlds  approaches. See, for example, the discussion by Bub in \cite{bub2020defense}.}, or a {\it classical-quantum  interpretation.}

Our main technical achievement is to show how both the Born rule and evolution laws of the quantum system are entirely determined by a change of measure, well-known for its use in probability theory and financial mathematics \cite{oksendal2003stochastic,bjork2009arbitrage}. In other words, the Born rule emerges from
classical probability theory applied to the classical system.
Mapping the bare noise in the classical system into one with a drift given by backreaction from the quantum system, we provide a class of quantum-like theories each parameterised by a specific choice of positive function on Hilbert space. In the case that this function is invariant under unitary transformations, we show that this necessarily leads to standard quantum theory. The alternatives appear to be worthy of study in their own right.

The dynamics we find are well known in the literature on classical-quantum dynamics. Since their early conception these theories have been appealing as providing an explicit model of quantum measurement \cite{blanchard1993interaction} and gravitational decoherence \cite{diosi1995quantum}. However, to date these theories have either been derived using quantum measurement \cite{HalliwelDH,kafri2014classical,2016Tilloy} or derived assuming %consistency with quantum theory
the quantum system is described by a density matrix, and required that the dynamics be completely-positive, which itself assumes the validity of the Born rule \cite{layton2024healthier,oppenheim2022two,oppenheim2023postquantum}. By showing that such theories may be derived axiomatically without requiring the measurement postulate of quantum theory, we rid these theories of the ontological baggage associated with quantum theory. Indeed, thinking of these theories in terms of continuous measurement, which is fed into an additional degree of freedom, is philosophically unnecessary – Occam's razor would require us to favour the current theory, in which no explicit statement of the Born rule is necessary. While the resulting quantum dynamics resembles that of spontaneous collapse models \cite{BassiCollapse,gisin1989stochastic}, the  collapse found here does not lead to decoherence of the quantum state, with the information instead encoded into a classical degree of freedom, such that the quantum state remains pure at all times. In this sense, the classical-quantum picture of quantum mechanics shares as much with features of the quantum darwinism and decoherence program \cite{schlosshauer2004decoherence}, with the classical degree of freedom acting as an environment that selects a preferred basis in the system.

%An obvious criticism of this approach is the problem of deciding which systems should be treated classically, and which quantum. We do not attempt to propose a universal criterion for this here. However, since this criterion does not need to come from quantum theory itself, solving this problem is strictly no harder than the measurement problem we have traded in for. One possible solution is to identify a fundamentally classical physical degree of freedom, an obvious candidate being spacetime \cite{oppenheim_post-quantum_2018}. However, one may equally take a pragmatic stance, identifying effective classicality in systems such as measuring apparatuses, the experimenter's notebook or Wigner's friend. The resulting theory is thus expected to applicable to a wide range of scenarios in which classicality, fundamental or effective, can be identified.

%An obvious question
%this approach raises, is 
%whether there is a system which should be treated classically.

An obvious feature of 
this approach is the problem of deciding which systems should be treated classically, and which quantum. We do not attempt to propose a universal criterion for this here. However, since this criterion does not need to come from quantum theory itself, solving this problem is strictly no harder than the measurement problem we have traded in for. 

One possible solution is to identify a \textit{fundamental} classical degree of freedom in nature. A motivated choice is to treat the gravitational field as this classical degree of freedom \cite{diosi2011gravity,kafri2014classical,2016Tilloy,PoulinKITP,oppenheim2023postquantum}. Aside from the difficulties of constructing  a quantum theory of gravity, this proposal is compelling because the central object of gravity is spacetime~\cite{oppenheim2023time}. Treating spacetime as a classical structure, may be a necessary requirement for quantum field theory to be well-posed. For example, the microcausality condition for fields
$
[\phi(x),\phi(y)]=0$\ for spacelike $x,y$
presupposes a definite (classical) causal structure with respect to which the notion of spacelike and timelike can be defined. Nor is gravity a gauge theory in the usual sense~\cite{notgauge}, which stands in contrast to the other forces for which we have found a quantum description. As an alternative to having a classical gravitational field, one could instead treat gravity as quantum, postulating the existence of a classical field that induces measurement-like dynamics \cite{Weller-Davies:2024zcb}. 

Whether spacetime is \emph{fundamentally} classical may be a less meaningful question than whether \emph{treating it} as classical is the appropriate description. Physics is merely the task of making better approximations of  our universe. Along these lines, we may instead motivate this interpretation of quantum theory using an \textit{effective} classical degree of freedom. Beyond the gravitational field, we may adopt an even more pragmatic stance,
identifying effective classicality in systems such as measuring apparatuses, the experimenter's notebook or Wigner's friend. This captures a notion of effective classicality that may  arise from a fully quantum theory \cite{layton2024classical}, or another theory entirely. In doing so, we extend the traditional notion of an effective theory beyond a simple separation of energy scales.
 
 %The resulting theory is thus expected to applicable to a wide range of scenarios in which classicality, fundamental or effective, can be identified.

\section{Axioms}
Let us first define a number of mathematical preliminaries. We will denote a Hilbert space by $\mathbb{C}^n$, linear operators on Hilbert space by capital letters $A,B,\ldots$ and the corresponding projective Hilbert space $P\mathbb{C}^n$. We will denote stochastic processes using subscript $t$, $\mathbb{E}$ to denote their expectations and $\xi_t$ to describe (the time-derivative of) a generic diffusion process that may be written as an Ito integral. %satisfying $\mathbb{E}[\xi_t\xi_{s}]=\delta(t-s)$
 \\
%When Ito integrals or functions of Ito processes appear, we will assume the standard continuity/differentiability/integrability properties to guarantee that the usual rules of Ito calculus may be used.  \\ 

We now state four axioms which appear natural to describe a simple classical system interacting with a quantum system.  \\

\noindent \textbf{Axiom 1:} The classical system occupies a definite point $Z_t\in\mathbb{R}$ in  configuration space at all times $t$; the  quantum system 
%at all times $t$ 
is described by a vector in Hilbert space $|\psi\rangle_t\in \mathbb{C}^n$. \\

\noindent \textbf{Axiom 2:} The classical system experiences a force from the quantum system which depends on the ray $\psi_t \in P\mathbb{C}^n $, and a noise process $\xi_t$ % It can include a stochastic contribution $\xi_t$ whose statistical properties are  time-independent. % and is generated by a white noise process 

\begin{equation}
\label{eq: true_classical}
    \frac{dZ_t}{dt}=f(\psi_t,Z_t) +  \xi_t .
\end{equation}

\noindent \textbf{Axiom 3:} The quantum system evolves linearly for fixed $Z_t$ under the same noise process $\xi_t$ i.e.
\begin{equation}
   \label{eq: true_quantum} \frac{d|\psi\rangle}{dt} =A(Z_t)|\psi\rangle_t + B(Z_t) |\psi\rangle_t \xi_t.
\end{equation}

\noindent \textbf{Axiom 4:} The classical probability distribution at time $t$ depends only on $|\psi\rangle_t$ and not past values.\\

%\noindent Axiom 1 says that our description assigns a definite classical state and quantum state at all times.  Axiom 2 describes the general form of Markovian diffusion process for a classical system, plus a force from the quantum system which is invariant under rescaling on the quantum state. It also states that the noise in the classical system is independent of that of the quantum one. The first part of Axiom 3 ensures a basic notion of linearity in Hilbert space, while the second part guarantees that there is no redundancy in the description of the theory from an the perspective of an observer of the classical system. Axiom 4 is a notion of ...\\

\noindent 
 Axiom 2 states that the force of the quantum system on the classical one obeys the same symmetry as the quantum dynamics (i.e. invariance under rescaling), and that the classical system experiences a stochastic noise $\xi_t$. Here $\xi_t$ is assumed to generate general diffusive evolution, a fairly mild assumption given the uniqueness of these processes under Markovianity and continuity (for more details see Appendix \ref{sec:genNoise}). 
%2 also states that the noise in the classical system is independent of that of the quantum one. 
The first part of Axiom 3 ensures a basic notion of linearity in Hilbert space, while the second  part is motivated by Occam's razor. Having only one noise source also guarantees that there is no redundancy in the description of the theory from the perspective of an observer of the classical system. \\

\noindent It will turn out that these postulates are sufficient to characterise a class of quantum-like theories by a single positive function $g: \mathbb{C}^n\rightarrow \mathbb{R}$. However, we shall see that these postulates are insufficient to rule out all alternatives to quantum theory, such as real amplitude quantum theory \cite{rennou2021quantum,ying2025whether}. For this general class of theories we derive Equations \eqref{eq: g_constraint} and \eqref{eq: force}, which govern how such quantum-like systems evolve and interact with a classical system. Next, we show that by requiring the theory to be independent of the choice of basis, we uniquely recover quantum theory.\\

%To do so, we shall make the additional postulate: \\

%\noindent \textbf{Postulate 5:} The classical probability distribution $P(z,t)$ depends only on $ \langle \psi |\psi\rangle_t $.\\

\noindent \textbf{Theorem 1:} The only theory obeying Axioms 1-4 with $g$ invariant under unitary transformations is either trivial, or recovers the predictions of standard quantum mechanics i.e. unitary evolution of the quantum state and collapse of the wave function with probabilities given by the Born rule.\\

\noindent Here trivial refers to a theory in which the state of the quantum system has no effect on the classical one. Note that our proof does not rely on the composition of Hilbert spaces under tensor products, nor do we attempt to address deriving this postulate of quantum theory.

\section{Change of measure} \label{sec: change_of_measure}

  \noindent Our main tool in what follows is to note that if our Axioms 2 and 3 hold, the true evolution of the classical system may be obtained by \textit{a change of measure} from the classical evolution under the noise alone. A standard and powerful tool in the theory of stochastic processes \cite{bjork2009arbitrage}, this allows a stochastic process with drift $X_t$ to be described by a driftless process $\widetilde{X}_t$ and a stochastic process $g_t$ that weights the paths to ensure the same statistics of observables (see Appendix \ref{app: change_of_meas} for more details). In our setting, it allows us to write the probability distribution corresponding to $Z_t$ in terms of $\widetilde{Z}_t=\int_0^t \xi_s ds$, the hypothetical classical evolution if subjected to noise alone, and a change of measure $g_t$. By Axiom 4 it follows that $g_t$ must be purely a function of the quantum state $|\psi\rangle_t$. In other words, there exists a function $g:\mathbb{C}^n\rightarrow \mathbb{R}$ such that
\begin{equation}
    \label{eq: q_on_c_effect}
    P(z,t)=\mathbb{E}[g(|\psi\rangle_t)\delta(z-\widetilde{Z}_t)]
\end{equation} where here $P(z,t)$ is the classical probability distribution. This is well defined only if $g(|\psi\rangle)\geq 0$ for all $|\psi\rangle$ and
\begin{equation} \label{eq: g_norm}
    \mathbb{E}[g(|\psi\rangle_t)]=1.
\end{equation} 
The first condition ensures all probabilities are non-negative, while the second is the normalization condition ensuring total probability is conserved. 
%This later property gives two things. First, 
For the quantum state $|\psi\rangle_t$, this fixes the initial normalisation i.e. $|\psi\rangle_0$ must satisfy $g(|\psi\rangle_0)=1$. 
For Eq. \eqref{eq: g_norm} to hold at all times, $g_t$ must be a martingale, and thus its drift must vanish identically \cite{bjork2009arbitrage}.
%These conditions are cornerstone of stochastic process 

To proceed, we opt to write $|\psi\rangle_t \in \mathbb{C}^n$ compactly as the complex vector $x_t$, with components $x_1 \ldots x_n$. For ease of presentation we will assume that $\xi_t$ is simply a white noise process, with the more general case discussed in Appendix \ref{sec:genNoise}. Using this notation, and provided the function $g$ is twice differentiable, we find via Ito's rule that
\begin{equation} \label{eq: g_constraint}
     \nabla g^T A x +x^\dag A^\dag (\nabla g)^*+ \frac{1}{2}x^T B^T S B x + \frac{1}{2} x^\dag B^\dag S^* B^* x^* + x^\dag B^\dag H B x=0, 
\end{equation} must hold for all $x\in \mathbb{C^n}$ from the martingale property of $g(|\psi\rangle_t)$. Here $\nabla g$ is a column vector, and $H$ and $S$ matrices, with entries defined as:
\begin{equation}
    (\nabla g)_i = \frac{\partial g}{\partial x_i} \quad \quad S_{ij}=\frac{\partial^2 g}{\partial x_i \partial x_j}\quad\quad  H_{ij}=\frac{\partial^2 g}{\partial x^*_i \partial x_j},
\end{equation} while the $v^*$, $v^T$ and $v^\dag$ refer to complex conjugation, transpose, and complex transpose respectively. For a given choice of $g$, this constrains the form of quantum evolution described by the linear operators $A$ and $B$. Taking the derivative of both sides of \eqref{eq: q_on_c_effect} and using Ito's rule, we find that the force on the classical system takes the form
\begin{equation} \label{eq: force}
f(x,x^*)=g^{-1}(\nabla g^T B x_t +x_t^\dag B^\dag (\nabla g)^*).
\end{equation} Thus the function $g$ determines (1) the normalised states of the theory (2) constrains the possible quantum evolution and (3) determines how the quantum system back-reacts on the classical one.

\section{Quantum mechanics as a classical-quantum theory} \label{sec: QM as CQ}

To see how the usual rules of quantum mechanics arises in this framework, we consider the special choice of  $g(x)=x^\dag x$.\\

\noindent We start by noting that the normalised states defined by $g(x)=1$ are given by the usual condition $x^\dag x=1$. To find the form of quantum dynamics and the force on the classical system, we compute $\nabla g$, $S$ and $H$, which take the form
\begin{equation}
    \nabla g = x^* \quad \quad S=0 \quad \quad H=I
\end{equation} where $I$ denotes the identity matrix. Inserting these into \eqref{eq: g_constraint} we find
\begin{equation}
    x^\dag (A + A^\dag + B^\dag B) x = 0
\end{equation} holds for all $x\in \mathbb{C}^n$ and thus that 
\begin{equation}
    A + A^\dag + B^\dag B = 0,
\end{equation} which has the unique solution $A=-iG + B^\dag B$ where $G$ here is a Hermitian operator. Substituting into \eqref{eq: true_quantum} we find quantum dynamics of the form
\begin{equation}
    \frac{d|\psi\rangle}{dt}=-iG|\psi\rangle  - \frac{1}{2}B^\dag B |\psi\rangle  + B |\psi\rangle \xi_t.
\end{equation} To find the dynamics for $Z_t$ we substitute into  \eqref{eq: true_classical} using \eqref{eq: force} to find
\begin{equation}
    \frac{dZ}{dt}=\frac{\langle \psi | (B + B^\dag) |\psi\rangle}{\langle \psi|\psi\rangle} + \xi_t.\\
\end{equation} In the above, both $B$ and $G$ may depend on the classical state $Z_t$.

For those familiar with continuous quantum measurement theory or spontaneous collapse models \cite{BassiCollapse,gisin1989stochastic, wiseman_milburn_2009,2006Jacob}, this dynamics will be immediately obvious as consistent with a general description of unitary dynamics and quantum measurement/decoherence. However, since the operators $G$ and $B$ are functions of the classical trajectory $Z_t$, the evolution of the quantum system, when considered in isolation, is neither Markovian or even linear. The dynamics are thus somewhat more akin to the examples of non-Markovian stochastic Schrödinger equations studied in ~\cite{diosi1998non,tilloy2021non}, although the resulting evolution equations on the quantum system, as well as the nature of the `environment', are significantly different.

For the unfamiliar, we note first that averaging over many realisations in the classical system reveals an average drift given by the first term, an expectation value of a Hermitian operator i.e. that the force in the classical system encodes the Born rule. The quantum dynamics reflects this evolution: when the force on the classical system is zero the quantum state evolves unitarily, while when $B$ is non-zero the additional terms acts to force the system into an eigenstate of the operator $B$ \cite{gisin1989stochastic}.

\section{Linear picture of quantum mechanics} \label{sec: linear}

While the above dynamics provides the true picture of evolution of classical and quantum systems, the change of measure allows us to study quantum mechanics in a non-interacting, entirely linear picture.

To see this, we note that normalising the quantum state dynamics (see e.g. \cite{2006Jacob}) one finds the evolution
\begin{equation}
dZ_t=\langle \psi | B(Z_t) + B^\dag(Z_t) |\psi\rangle_t dt + dW_t
\end{equation}

\begin{equation}
    d|\psi\rangle_t = A(Z_t) |\psi\rangle dt + B(Z_t) |\psi\rangle dW_t
\end{equation} which 
is equivalent to the unnormalised dynamics
\begin{equation} \label{eq: linear_C}
    d\widetilde{Z}_t=dW_t
\end{equation}

\begin{equation} \label{eq: linear_Q}
    d|\widetilde{\psi}\rangle_t = A(\widetilde{Z}_t) |\widetilde{\psi}\rangle dt + B(\widetilde{Z}_t) |\widetilde{\psi}\rangle dW_t
\end{equation}  In the first picture we have the true evolution of the classical and quantum systems, with the state of the classical system affecting the quantum one and vice versa. In the second picture, both systems evolve purely under the noise process (here simply a white noise process), with no direct interactions between the two. Here we have explicitly included a tilde over the quantum state in the second set of dynamics, to emphasise that under the measure in this space, the evolution law depends on $\widetilde{Z}_t$ rather than $Z_t$. The two pictures are generally related by the equality 
\begin{equation}
    \mathbb{E}[|\psi\rangle_t \langle \psi |_t \delta (z-Z_t)]=\mathbb{E}[|\widetilde{\psi}\rangle_t \langle \widetilde{\psi} |_t \delta (z-\widetilde{Z}_t)]
\end{equation} which is a statement that the classical-quantum state is the same computed in either picture \cite{layton2024healthier}. Taking integrals or the trace on the left hand side allows the computation of general classical-quantum observables \cite{oppenheim2023postquantum,,weller2024classical} using the quantities $|\widetilde{\psi}\rangle_t$, $\widetilde{Z}_t$, demonstrating the full equivalence of the two descriptions.

\section{Beyond quantum theory?}

Not every classical-quantum theory obeying Axioms 1-4 takes the form of quantum mechanics. To see this, we consider new theories generated by alternative functions $g(x,x^*)$.\\

\noindent\textbf{Example 1: $g(x)=(x^\dag x)^2$}\\

\noindent Proceeding as in Section \ref{sec: QM as CQ} we find
\begin{equation}
    \nabla g = 2(x^\dag x)x^* \quad \quad S=2x^* x^\dag \quad \quad H=2x x^\dag + 2(x^\dag x)I
\end{equation} where here $x^* x^\dag$ and $x x^\dag$ are interpreted as outer products. This gives the condition \eqref{eq: g_constraint} as
\begin{equation}
    2(x^\dag x) x^\dag (A + A^\dag + B^\dag B) x + \big(x^\dag (B + B^\dag) x\big)^2 = 0
\end{equation}  holds for all $x\in \mathbb{C}^n$ which by the lemma in the next section implies $B+B^\dag=2bI$ where $b$ is a real number. Proceding as before we find that
\begin{equation}
\begin{split}
    f(x^*,x)= \frac{2 (x^\dag x) x^\dag (B + B^\dag) x}{(x^\dag x)^2} 
    = 4 b 
\end{split} 
\end{equation} i.e. recovering purely classical evolution when $b=b(Z_t)$, with no effect from the quantum system.\\

\noindent \textbf{Example 2:} $g(x)=(x+x^*)^\dag (x+x^*)$\\

\noindent Immediately, we see that for $g(x)=1$ to be satisfied, the initial states must be chosen to be real-valued. Again, we may compute the relevant quantities to find
\begin{equation}
    \nabla g = 2(x+x^\dag ) \quad \quad S=2I \quad \quad H=2I
\end{equation} which after some rearranging gives the condition \eqref{eq: g_constraint} as
\begin{equation}
    x^\dag(A + A^\dag + B^\dag B) x +  \mathrm{Re} \big\{x^T (A + A^T + B^T B) x\big\}= 0
\end{equation}  which holds for all $x\in \mathbb{C}^n$. Using the polarisation identity it is straightforward to show that this implies the two conditions
\begin{equation}
   A+ A^\dag + B^\dag B=0 
\end{equation}
\begin{equation}
  \mathrm{Re} \big\{A + A^T + B^T B\big\}=0,
\end{equation} which may be rearranged to show that
\begin{equation}
    B=B^*,
\end{equation} i.e. that the matrix $B$ must be real. Computing now the dynamics on $Z$ using \eqref{eq: force} we find the force is given
\begin{equation}
f(x,x^*)=\frac{(x+x^*)^\dag 2 B (x+ x^*)}{(x+x^*)^\dag (x+x^*)} .
\end{equation} We thus see that we have recovered the Born rule of real amplitude quantum mechanics, an alternative theory in which all the vectors and operators are taken to be real valued \cite{rennou2021quantum,ying2025whether}.\\

\noindent\textbf{Example 3: $g(x)=x^\dag T x$}\\

\noindent Here the normalised states are given $x^\dag T x = 1$ – one may check for $\mathbb{C}^2$ that this defines a ``Bloch ellipsoid", rather than a Bloch sphere.  To study the quantum dynamics and back-reaction in this theory, we compute $\nabla g$, $S$ and $H$, which take the form
\begin{equation}
    \nabla g = T^T x^*  \quad \quad S=0 \quad \quad H=T
\end{equation} where $I$ denotes the identity matrix. Since $T$ is constant and $g(x)\geq 0$ it implies $T\succeq 0$ i.e. that $T$ is positive semi-definite. Inserting these into \eqref{eq: g_constraint} we find
\begin{equation}
    x^\dag (T A + A^\dag T + B^\dag T B) x = 0
\end{equation} holds for all $x\in \mathbb{C}^n$ and thus that 
\begin{equation}
    TA + A^\dag T + B^\dag T B = 0. 
\end{equation} We find this has a solution for $A$ of the form 
\begin{equation}
    A=-i T^{-1} G - \frac{1}{2}T^{-1}B^\dag T B
\end{equation} where $G$ is a Hermitian operator i.e. unless $[T^{-1},G]=0$ we see that the quantum dynamics do not recover the usual anti-Hermitian term that we identify with the $-i$ times the Hamiltonian.  \\

\noindent To find the quantum back-reaction on the classical evolution we substitute $\nabla g$, $S$ and $H$, into \eqref{eq: force} to find
\begin{equation}
    f(x,x^*)=\frac{x_t^\dag (T B + B^\dag T) x_t}{x_t^\dag T x_t } 
\end{equation} which may be rewritten as
\begin{equation}
f(x,x^*)=\frac{x_t^\dag T^{\frac{1}{2}} (T^{\frac{1}{2}} B T^{-\frac{1}{2}} + T^{-\frac{1}{2}} B^\dag T^{\frac{1}{2}}) T^{\frac{1}{2}}x_t}{x_t^\dag T^{\frac{1}{2}} T^{\frac{1}{2}} x_t }
\end{equation} i.e. we again find a force based on the expectation value of a Hermitian operator.\\

\section{Proof of Theorem 1} \label{sec: theorem}

\noindent Here we let $g$ be an arbitrary positive function $h$ of the norm squared $x^\dag x$ – this is sufficient since any function that is unitarily invariant may be expressed as a function of the norm. It is straightforward to compute that
\begin{equation}
    \nabla g = h^\prime x^*  \quad \quad S=h^{\prime \prime} x^* x^\dag \quad \quad H=h^{\prime \prime} x^* x^\dag + h^\prime I
\end{equation} where $I$ denotes the identity matrix. Plugging into \eqref{eq: g_constraint} and rearranging we find the condition
\begin{equation}
\label{eq: h_condition}
    h^\prime \big[x^\dag (A + A^\dag + B^\dag B) x \big] + \frac{h^{\prime \prime}}{2} \big[x^\dag (B + B^\dag ) x\big]^2 = 0.
\end{equation} To proceed we shall make use of the following lemma. \\

\noindent \textbf{Lemma:} If operators $N$ and $M$ satisfy
\begin{equation} \label{eq: MN_lemma_eq}
     x^\dag M x = \big(x^\dag N x\big)^2
\end{equation} for all $||x||=1$, then  $M=n^2 \mathds{1}$ and $N=n\mathds{1}$ for $n\in  \mathbb{R}$. \\

\noindent To prove this, we first note that we can rewrite this using vectors with arbitrary norm as
\begin{equation} \label{eq: MandN_arb_norm}
    x^\dag M x x^\dag x = \big(x^\dag N x\big)^2,
\end{equation} and then expand in the basis as
\begin{equation}
    \sum_{ijkl}x_i^* x_j^* x_k x_l [M_{ik}\delta_{jl}-N_{ik}N_{jl}]=0.
\end{equation} Taking successive derivatives with respect to $x$ and $x^*$ one finds that this complex polynomial is identically zero for all $x$ only if the symmetrised sum of the coefficients vanishes i.e. if
\begin{equation}
    M_{ik}\delta_{jl} + M_{jk}\delta_{il} + M_{il}\delta_{jk} + M_{jl}\delta_{ik}=2(N_{ik} N_{jl}+N_{jk} N_{il})
\end{equation} Setting $i=j=k=l$ and summing over $l$ gives 
\begin{equation}
\textrm{tr}(M)=\textrm{tr}(N^2),
\end{equation}while setting $i=l$ and summing over $l$ gives \begin{equation}
    2M + \textrm{tr}(I) M  + \textrm{tr}(M) I = 2N^2 + 2 \textrm{tr}(N)N,
\end{equation} and hence
\begin{equation}
    (1+ \textrm{tr}(I))\textrm{tr}(M) = \textrm{tr}(N^2)+ \textrm{tr}(N)^2.
\end{equation}Combining these two expressions for $\textrm{tr}(M)$ we find
\begin{equation}
    \textrm{tr}(N)^2=\textrm{tr}(N^2)\textrm{tr}(I)
\end{equation} which saturates the Cauchy-Schwartz inequality for the inner product $\textrm{tr}(AB)$, and thus implies that
$N$ and $I$ must be linearly dependent i.e. $N=n\mathds{1}$ for $n\in \mathbb{R}$. Substitution of this result for $N$ readily provides the corresponding expression for $M$.\\

\noindent Returning to \eqref{eq: h_condition}, we use this lemma to rule out the trivial case where $h^{\prime \prime}$ is non-zero. In this case, we may rearrange the expression into the form of \eqref{eq: MN_lemma_eq}
\begin{equation}
    M=-\frac{2h^\prime}{h^{\prime \prime}}(A+A^\dag + B^\dag B) \quad \quad  N=B+B^\dag
\end{equation} from which it follows that  $B+B^\dag$ is proportional to the identity i.e. $B+B^\dag = n \mathds{1}$. Substituting into the force on the classical system \eqref{eq: true_classical} we see that
\begin{equation}
f(x,x^*)=\frac{h^\prime(x^\dag x) (x^\dag (B+B^\dag) x)}{h(x^\dag x)}=\frac{n h^\prime(x^\dag x) x^\dag x}{h(x^\dag x)}
\end{equation} i.e. the force purely a function of the norm. However, by Axiom 2, the force on the classical system is a function of the ray and not the ray representative, and thus must be invariant under rescaling. Thus the force must be a constant. The state of the quantum system thus has no effect on the classical one and thus describes a trivial theory. \\

\noindent The other and non-trivial possibility is that $h^{\prime \prime}=0$. Since the second derivative of $h$ vanishes for all $x$, it follows that \begin{equation}
    h(x^\dag x)= c x^\dag x + c_0.
\end{equation}Substituting this solution into the force on the classical system we see that
\begin{equation}
f(x,x^*)=\frac{ cx^\dag (B+B^\dag) x}{c x^\dag x + c_0}
\end{equation} which to be invariant under rescaling of the state vector as required by Axiom 2 must imply that $c_0$ is zero. We thus recover $g(x,x^*)= c x^\dag x$, where $c$ is an arbitrary constant that we can set to 1 if we choose to normalise the inital states by $x^\dag x = 1$. The dynamics determined by $g(x,x^*)=x^\dag x$ is exactly that studied in Sec. \ref{sec: QM as CQ} and thus is shown to be equivalent to the standard predictions of quantum theory.

\section{Comparison to other interpretations}

% \noindent \textbf{Stochastic collapse:} Here, the field which is causing the collapse is the classical field, and can be back-reacted on. It's role is not to cause collapse according to the Born role, but it has this effect (ontologically simpler). We  cannot think of the mechanism as the quantum system being 'hit' or weakly measured, rather there is a complicated non-Markovian interaction between the classical and quantum field.   Need to understand renormalisation better to understand if there is a problem of the tails (divergences at shorter distances) of the kind one has in stochastic collapse models. \\
\noindent \textbf{Stochastic collapse:} Traditional stochastic collapse models \cite{Ghirardi1986GRW, Pearle_2012, BassiCollapse, Bassi2003CollapseReview}, and related gravitationally induced collapse proposals like the Diosi–Penrose (DP) model \cite{Penrose1996,diosi2011gravity}, share a common structure: they postulate a nonlinear stochastic modification to the Schrödinger equation acting directly on the quantum state. This modification induces effective collapse dynamics, often chosen to reproduce the Born rule. However, these approaches typically do not include a dynamical model for the degrees of freedom responsible for collapse. The stochastic field is introduced phenomenologically as an external noise source, and while it affects the quantum state, there is no corresponding back-reaction onto the field itself. 

In the classical-quantum framework, by contrast, a pre-existing classical field (such as spacetime curvature) necessarily causes collapse simply by being coupled self-consistently to the quantum system, as has been noted before \cite{hall2005interacting,2016Tilloy, poulinPC}. This classical field can be back-reacted upon, ensuring a consistent exchange of information between the quantum and classical sectors. Only if the classical field is ignored is their decoherence. The collapse of the quantum state is not postulated but emerges from this mutual, non-Markovian interaction. Ontologically, this is simpler and more symmetric: both sectors evolve under the same coupled dynamics, rather than one acting unilaterally on the other. Further, because the coupling is non-Markovian, the classical field retains memory of its past interactions, meaning that conditioned on the classical degree of freedom the quantum state remains pure and there is no fundamental decoherence conditioned on the classical state. 

There is however a danger that the classical-quantum theory inherits an undesirable feature of stochastic models -- namely that they lead to anomalous heating which is tightly constrained experimentally  \cite{donadi2021underground}. However, this is a short distance effect \cite{bps}, 
and a direct application of that analysis is not straightforward. At short distances, the interaction between a quantum field and a classical field is expected to be highly non-Markovian, and the non-linearities in the coupled dynamics become significant. A full treatment would require a renormalization of the theory, a procedure which is not yet fully understood for hybrid dynamics. Encouragingly, recent work on the renormalization of post-quantum gravity suggests that such theories can be well-behaved, with a dynamical length scale emerging akin to that in QCD \cite{grudka2024renormalisation}. Nevertheless, whether this mechanism is sufficient to regulate the dynamics and evade heating constraints is a crucial question that must be more thoroughly studied.
\\

\noindent \textbf{Weak-measurement and feedback: }% and post-quantum:} 
Our representation connects to the class of continuous measurement and feedback formulations \cite{2016Tilloy,layton2024healthier,Tilloy:2024qpi}, which show that certain stochastic collapse equations can be recast as feedback-controlled measurement dynamics. In the usual derivation of such equations, one begins with the Born rule and notion of measurement, and postulates that the system is weakly and continuously measured by an external classical apparatus in order to arrive at self-consistent equations. Once averaged over the classical trajectories, such dynamics is equivalent to linear evolution on the CQ state \cite{layton2024healthier, Tilloy:2024qpi} that has been introduced in post-quantum theories of gravity \cite{oppenheim_post-quantum_2018}. 

Care should be taken because at short distances, the interaction of a quantum field and a classical field is highly non-Markovian. Nonetheless, if the feedback from the classical field is updated arbitrarily quickly, one arrives at a classical-quantum interpretation by promoting the classical degree of freedom that acts as a measurement record to a dynamical degree of freedom. Conversely, in our setting, one postulates a dynamical classical degree of freedom from the start, and the weak measurement equations emerge dynamically from the coupling between a classical field and a quantum system, without invoking the Born rule or the notion of measurements as an axiom.   \\

% While formally equivalent, as shown in \cite{layton2024healthier,tilloy}, Not really an interpretation but...\\

\noindent \textbf{Many-worlds / Quantum Darwinism:} 
The structure of our theory—a pure quantum state conditioned on a classical record—bears a resemblance to frameworks like Quantum Darwinism and, by extension, the Many-Worlds Interpretation \cite{Zurek2003Decoherence,Zurek2009QuantumDarwinism, riedel2016objective}. In those approaches, the ``classical record" is not a fundamental entity but is identified with a redundant fragment of the quantum environment. The total state of the system and environment remains globally pure, while the state of the system conditioned on a specific state of the environmental fragment appears to collapse, mimicking a measurement record.

Every possible state of the environment is realised in a different branch of the universal wavefunction. The difficulty is how to assign weights to each of these co-existing branches, which usually require additional assumptions \cite{Wallace2012EmergentMultiverse,hardy2001quantum,Zurek2003Decoherence,zurek2005probabilities}.   Furthermore, an observer who traces out the environment sees a mixed density matrix and how to interpret this is unclear. Proponents argue that from within any given branch, the world appears to be a single, definite reality.

By contrast, here, only a single, definite trajectory ever occurs. Density matrices and probability distributions thus arise only from classical, subjective ignorance of the true state of the classical system. This places the foundation of probability in our theory on a much firmer and more conventional footing: probability is simply a measure of our lack of knowledge about the single, ontic, but stochastic, state of the world, a picture entirely in line with a standard Bayesian or classical statistical viewpoint.
\\

\noindent \textbf{$\psi$-epistemic} Our construction may also be interpreted in terms of $\psi$-epistemic statements. The assumption in Axiom 3 that the noise appearing in both the classical and quantum systems is the same amounts to requiring that the quantum state is an object uniquely determined by observations of the classical system. In this setting, Axiom 4 states that the quantum state contains all the information required to update the classical system, similarly to how a conditional probability distribution behaves in classical filtering theory \cite{zakai1969optimal}, as we show in Appendix \ref{app: filtering}.\\

\noindent \textbf{De Broglie-Bohm:} The interpretation of the quantum state as defining a measure bears a conceptual similarity to the De Broglie-Bohm, or pilot-wave, formulation of quantum mechanics \cite{Bohm1952Interpretation,Durr2009Bohmian}. In the pilot wave picture, the wavefunction acts a guide for a set of hidden variables (the definite particle positions $r_i$) that evolve deterministically under a force proportional to the local conditional momentum expectation, $f(\psi_t) \sim \frac{\langle \psi| p + p^{\dag} |\psi \rangle}{\langle \psi | \psi \rangle}$, where $p$ is the momentum operator. The wavefunction evolves according to unitary dynamics and there is no back-reaction from the particle configuration $r_i$. The stochasticity in pilot wave theory arises due to a measure $g(|\psi\rangle_t) = \langle \psi |\psi \rangle_t$ that is purely epistemic, reflecting ignorance of the initial conditions rather than any underlying ontic noise process. In contrast, the classical system in our framework is not representing a hidden variable for the quantum state but rather a fundamental (or effectively classical) degree of freedom. The classical and quantum variables evolve under a coupled stochastic process that allows for mutual back-reaction. In this sense, the pilot wave theory violates our Axioms 2-3 since it does not undergo a coupled noise process, even though it exhibits structural parallels with the formulation of quantum mechanics as a classical-quantum theory in Section \ref{sec: QM as CQ}, both through the pilot wave force and the measure. \\

\noindent\textbf{Comparison to other derivations of the Born rule:} Our derivation of the Born rule from axioms follows a long line of such efforts \cite{Gleason1957Measures,busch2003quantum,Wallace2012EmergentMultiverse,hardy2001quantum,Zurek2003Decoherence,zurek2005probabilities,barrett2007information,masanes2019measurement,hossenfelder2021derivation}. Our approach shares many similarities with that of Gleason, in the sense that both approaches involve assigning probability distributions to pure states in Hilbert space \cite{Gleason1957Measures}. However, while in Gleason's case these are assumed orthogonal (or more generally, of a POVM form \cite{busch2003quantum}), here the outcomes are entirely determined by dynamics, which are not fixed a priori. More generally, our measurement free, dynamical approach is even more in contrast to more modern approaches to deriving the Born rule \cite{hardy2001quantum,barrett2007information,masanes2019measurement} which often employ a notion of composition of systems and partial trace, which are strong assumptions.

\section{Discussion}

We have argued here for a basic addition to standard quantum theory – namely that of a classical system. A key assumption of this was to identify noise as inherent in both the classical system and quantum system's evolution. Since noise typically arises from tracing out an environment, it is tempting to imagine that such a description could arise from a fully quantum theory. While this is known to be true in certain cases \cite{layton2024classical}, it may not true in general, at least for all times. The essential reason is that classically simulating non-commuting observables while coupling them to a quantum system may not be possible for sufficiently long times. The notion of complete positivity is weaker for classical systems in comparison to quantum ones. However, we have no proof in either direction, and this therefore remains an interesting open question. 

%Baked into our assumptions were that of a Markovian representation for the noise $\xi_t$ in the system. It would be interesting exploring the current work in the context of non-Markovian unravellings \cite{diosi1998non}, of which some connections have already been made to interpretations of quantum theory \cite{tilloy2021non}

A key technical result of this work was to show a change of measure may be used to represent the coupled dynamics of quantum systems in an alternative picture where the quantum state evolution is entirely linear, and the classical state evolves without back-reaction from the quantum one. This linear and non-interacting representation of quantum mechanics has important practical outcomes: namely providing a method for studying measurement and feedback that decouples the measurement signal and quantum system \cite{layton2025linear}, and more generally a promising approach to studying relativistic classical-quantum dynamics where the back-reaction on the metric is encoded in the norm of the quantum state, rather than explicitly via back-reaction term as in the semi-classical Einstein equations. 

Finally, we note that a key axiom of quantum theory is the composition of subsystems under the tensor product, which we did not attempt to derive here. Understanding whether it is possible to derive this axiom in our framework, and understanding how non-locality interplays with the linear, non-interacting picture of quantum theory we have introduced, is an interesting problem that we leave to future work.
\\

\noindent
\textbf{Acknowledgements:} We would like to graciously thank Maite Arcos, Todd Brun, Lluis Masanes, Emanuele Panella, Jess Riedel, Andrea Russo for helpful and stimulating discussions.

%\bibliography{library}
%\bibliographystyle{unsrt}
\printbibliography
\newpage

\appendix

\section{General noise} \label{sec:genNoise} 

To make the following discussion precise in this appendix (and the remaining appendices), we use explicitly the Ito formalism, namely the Wiener increment $dW_t$ satisfying $dW_t^2=dt$, rather than $\xi_t$ as in the main text. This gives the main dynamics we consider as
\begin{equation}
    dZ_t=f(\psi_t,Z_t) +  dW_t .
\end{equation} and
\begin{equation}
d|\psi\rangle=A(Z_t)|\psi\rangle_t + B(Z_t) |\psi\rangle_t dW_t,
\end{equation} which can be thought of as the more precise statements of \eqref{eq: true_classical} and \eqref{eq: true_quantum} under the basic white noise case $\mathbb{E}[\xi_t \xi_s]=\delta(t-s)$.

Having stated this case clearly, we now review how our results extend to the more general case where the noise $\xi_t$ in the classical system may be dependent on the classical trajectory.

First, note that if $\xi_t$ is the time derivative of a diffusive Ito process $R_t$, then $R_t=\int_0^t \xi_s$ generically evolves under
\begin{equation} \label{eq: gen_noise}
dR_t=\mu(\omega,t) dt + \sigma(\omega,t) dW_t
\end{equation} where here $\mu(\omega,t)$ and $\sigma(\omega,t)$ are any functions that are determined given the Wiener process $W_s$ up to time $t$ (i.e. adapted) \cite{oksendal2003stochastic}. This dynamics includes the case where $\mu$, $\sigma$ are explicit functions of the $Z_t$, as well as if they are functionals of this process. If $\mu(\omega,t)$ is non-zero, we may simply redefine $f$ in Eq. \eqref{eq: true_classical} and $A$ in Eq. \eqref{eq: true_quantum} to arrive at a process that is drift free. This guarantees that we may always exploit the change of measure trick used in Appendix \ref{app: change_of_meas}. To study the case that $\sigma(\omega,t)$ is not equal to 1 i.e. where $\xi_t$ is not white noise, we note that the resulting computations lead to the force $f$ being multiplied by a factor $\sigma$, and the equations for the quantum state being redefined by $B\mapsto \sigma B$. In the limit that $\sigma$ vanishes, we thus see that both $f$ and $B$ vanish i.e. removing the non-unitary part of the quantum evolution.

Secondly, we note that the assumption that $\xi_t$ may be defined in the above form is rather mild. Aside from being the basis of modelling a wide range of continuous noise processes in nature \cite{oksendal2003stochastic}, it also describes any process that is defined in terms of a process that is continuous, has independent increments and is stationary. This is guaranteed by the Levy-Ito decomposition theorem which states that any stationary process with stationary increments can be uniquely decomposed into a drift, a continuous part, and a jump part. For the trajectory to be continuous, the jump component must be zero, leaving only the continuous part, which is uniquely described by a Wiener process. A broader class of processes are those defined in terms of the Wiener process, i.e. those given by \eqref{eq: gen_noise}, which is the class of noise processes we consider.

\section{Change of measure} \label{app: change_of_meas}A change of measure provides a method for changing the drift in a stochastic differential equation \cite{oksendal2003stochastic,bjork2009arbitrage}. Consider the following stochastic process
\begin{equation}
    dX_t = \mu_t dt + dW_t.
\end{equation} Then provided the drift $\mu_t$ is adapted to the Wiener measure i.e. deducible at time $t$ from observations of $W_s$ up to time $s \leq t$, then there exists a change of measure process (or likelihood process) $g_t$ such that another driftless process
\begin{equation}
    d\tilde{X}_t = dW_t
\end{equation} satisfies
\begin{equation}
    \mathbb{E}[a(X_t)]=\mathbb{E}[g_t a(\widetilde{X}_t)],
\end{equation} for any function $a$. The adaptedness of the drift indeed holds in our setting, since setting $\mu_t=f(|\psi\rangle_t, Z_t)$, one sees from Axiom 2 and 3 is the only noise process that appears and appears everywhere time-locally. 

For our purposes the above statement is sufficient, since we make use of Axiom 4, and require the necessary properties of $g_t$ using its functional dependence on $|\psi\rangle_t$. However, Girsanov's theorem also provides an explicit realisation of this change of measure $g_t$ in terms of the drift $\mu_t$, which takes the form
\begin{equation}
    g_t=e^{\int_0^t \mu_s dW_s - \frac{1}{2}\int_0^t \mu_s^2 ds}.
\end{equation} In this case the function is required to be suitably regular (c.f. the Novikov condition \cite{bjork2009arbitrage}) to be a well-defined martingale and hence satisfy $\mathbb{E}[g_t]=1$. Note that the time non-local dependence on the drift in this $g_t$ provides some explanation of how Axiom 4 narrows down the possible forces $f$ on the classical system to the specific form given in Eq. \eqref{eq: force}.

\section{$g$-derived quantities} \label{g_quantities}

\noindent This lets us write the main dynamics as
%\begin{equation}
%    
%\end{equation}
\begin{equation}
    |\psi\rangle = \sum_i x_i |i\rangle
\end{equation} and use $x=(x_1,\ldots,x_n)^T$ to denote the column vector associated to this – this provides a convenient representation of the quantum state. The dynamics can then be written as
\begin{equation}
    dx_i=a_{ij} x_j dt + b_{ij} x_j dW_t,
\end{equation} or as
\begin{equation}
    dx_t=A x_t dt + B x_t dW_t,
\end{equation} where $A$ and $B$ are taken to be the matrix representations of the operators. Writing $g(|\psi\rangle)=g(x)$, we find from Eq. \eqref{eq: g_norm} that 
\begin{equation}
    \mathbb{E}[dg(x_t)]=0.
\end{equation} It is a straightforward calculation using Ito's rule to check that
\begin{equation}
    dg=\nabla g^T dx_t +dx_t^\dag  (\nabla g)^* dt 
    + \frac{1}{2}dx_t^T S_t dx_t + \frac{1}{2} dx_t^\dag  S_t^* dx_t^* +dx_t^\dag  H_t dx_t,
\end{equation} and thus
\begin{equation}
\begin{split}
    dg=&\nabla g^T A x_t dt +x_t^\dag A^\dag (\nabla g)^* dt + \nabla g^T B x_t dW_t +x_t^\dag B^\dag (\nabla g)^* dW_t \\
    &+ \frac{1}{2}x_t^T B^T S_t B x_t dt + \frac{1}{2} x_t^\dag B^\dag S_t^* B^* x_t^* dt+x_t^\dag B^\dag H_t B x_t dt.
\end{split}
\end{equation}This implies
\begin{equation}
    \mathbb{E}[\nabla g^T A x_t +x_t^\dag A^\dag (\nabla g)^* + \frac{1}{2}x_t^T B^T S_t B x_t + \frac{1}{2} x_t^\dag B^\dag S_t^* B^* x_t^*+x_t^\dag B^\dag H_t B x_t ]=0,
\end{equation} which by the Martingale property \cite{bjork2009arbitrage} implies
\begin{equation} \label{eq: g_constraint_app}
     \nabla g^T A x +x^\dag A^\dag (\nabla g)^*+ \frac{1}{2}x^T B^T S B x + \frac{1}{2} x^\dag B^\dag S^* B^* x^* + x^\dag B^\dag H B x=0 
\end{equation} holds for all $x\in \mathbb{C}^n$. Here $\nabla g$ is a column vector, and $H$ and $S$ matrices, with entries defined as:
\begin{equation}
    (\nabla g)_i = \frac{\partial g}{\partial x_i} \quad \quad S_{ij}=\frac{\partial^2 g}{\partial x_i \partial x_j}\quad\quad  H_{ij}=\frac{\partial^2 g}{\partial x^*_i \partial x_j},
\end{equation} while the $v^*$, $v^T$ and $v^\dag$ refer to complex conjugation, transpose, and complex transpose respectively.\\

\noindent To study the effect on the true classical evolution, we note that we can consider the time evolution of \eqref{eq: q_on_c_effect} to find a useful expression. First, we  write
\begin{equation}
    \frac{\partial P(z)}{\partial t}=\mathbb{E}[dg \delta(z-\tilde{Z}_t) + g d\delta(z-\tilde{Z}_t) + dg d\delta(z-\tilde{Z}_t)],
\end{equation} The first term on the right hand side vanishes by virtue of \eqref{eq: g_constraint_app} and $\mathbb{E}[f(x_t)dW_t]=0$ for Ito processes. Computing the second and third terms explicitly we find
\begin{equation}
    \mathbb{E}[g(x_t) d\delta(z-\tilde{Z}_t)]=\frac{1}{2}\frac{\partial^2 }{\partial z^2}\mathbb{E}[ g(x_t)\delta (z-\tilde{Z}_t) ] dt
\end{equation} 
\begin{equation}
    \mathbb{E}[dg(x_t) d\delta(z-\tilde{Z}_t)]=-\frac{\partial}{\partial z}\mathbb{E}[  (\nabla g^T B x_t +x_t^\dag B^\dag (\nabla g)^*) \delta (z-\tilde{Z}_t) ] dt.
\end{equation}
Comparing now to the expansion
\begin{equation}
\begin{split}
    \frac{\partial P(z)}{\partial t}=&\ \mathbb{E}[d\delta(z-Z_t)]\\
    =&-\frac{\partial}{\partial z}\mathbb{E}[\delta(z-Z_t) dZ_t] + \frac{1}{2}\frac{\partial^2}{\partial z^2}\mathbb{E}[\delta(z-Z_t) dZ_t^2] 
\end{split}
\end{equation} it is straightforward to see that the two expressions are equivalent for
\begin{equation}
    dZ_t = d\tilde{Z}+ g^{-1} dg d\tilde {Z}
\end{equation} i.e.
\begin{equation} \label{eq: true_classical_app}
    dZ_t=g^{-1}(\nabla g^T B x_t +x_t^\dag B^\dag (\nabla g)^*) dt + dW,
\end{equation}  
describes the true classical  evolution.\\

\section{The Zakai equation} \label{app: filtering}

We note here a formal similarity between our dynamics and that of classical filtering theory \cite{zakai1969optimal}. Consider a degree of freedom $Y_t$ that is unable to observed directly, but instead gives a response on an observed degree of freedom $Z_t$, according to the equations
\begin{equation}
    dY_t=h(Y_t) dt + dV_t
\end{equation}
\begin{equation} \label{eq: Z_class_filt}
    dZ_t=f(Y_t) dt + dW_t.
\end{equation} Then it is well known that the conditional probability distribution $P_t=P(y)_t$ given the observations of $Z_t$ can be described instead using the linear equation
\begin{equation} \label{eq: P_class_filt}
    dP_t(y)=L(P_t(y)) dt + f(y) P_t(y) dZ 
\end{equation} which is known as the Zakai equation, where 
\begin{equation}
    L(P)=-\frac{\partial}{\partial y} ( h P) + \frac{1}{2}\frac{\partial^2 P}{\partial y^2}.
\end{equation} To see the analogy with our dynamics, we transform to a non-interacting picture
\begin{equation}
    d\tilde{Z}_t = dW_t
\end{equation}
\begin{equation}
    d\tilde{P}_t(y) = L(\tilde{P}_t(y)) dt + f(y) \tilde{P}_t(y) dW_t
\end{equation} which is the analogue of \eqref{eq: linear_C} and \eqref{eq: linear_Q}. We can identify the change of measure with
\begin{equation}
    g_t = \int_{-\infty}^\infty dy \tilde{P}_t(y)
\end{equation} which we can straightforwardly check satisfies $\mathbb{E}[dg_t]=0$ and $g_0=1$ for $P_0(y)$ normalised. Recomputing the dynamics of $Z_t$ using this change of measure we find
\begin{equation}
    dZ_t=\frac{\int_{-\infty}^\infty dy f(y) P_t(y)}{\int_{-\infty}^\infty dy P_t(y)} dt + dW_t
\end{equation} which gives an entirely different expression for the drift in the system, that nevertheless guarantees the correct joint dynamics. To see this, we finally compute the dynamics of the quantity
\begin{equation}
    P(y,z,t)=\mathbb{E}[\tilde{P}_t(y)\delta(z-\tilde{Z}_t)]
\end{equation} using $\tilde{Z}_t$ and $\tilde{P}_t$, to find that it is given
\begin{equation}
    \frac{\partial P(y,z,t)}{\partial t}=-\frac{\partial}{\partial y} ( h P) + \frac{1}{2}\frac{\partial^2 P}{\partial y^2} -\frac{\partial}{\partial z} ( f P) +\frac{1}{2}\frac{\partial^2 P}{\partial z^2}
\end{equation} where the third term arises from the cross term in Ito's rule. The same dynamics are found for
\begin{equation}
    P(y,z,t)=\mathbb{E}[\delta(y-Y_t)\delta(z-Z_t)]
\end{equation} using either the form given in terms of $P_t$ or $Y_t$. We thus see that the correct joint dynamics are recovered, even in this non-interacting picture.

\end{document}